\documentclass[preprint,endfloats,pre,draft]{revtex4}
\usepackage{bm}
\usepackage[final]{graphicx}
\usepackage{longtable}
\usepackage{amsmath}
\newcommand{\be}{\begin{equation}}
\newcommand{\ee}{\end{equation}}
\newcommand{\bea}{\begin{eqnarray}}
\newcommand{\eea}{\end{eqnarray}}

\begin{document}

\title{Influence of long-range correlated surface and near the surface disorder
on the process of adsorption of long-flexible polymer chains.}

\author{Z.Usatenko$^{1}$,$^{2}$,$^{3}$, J.-U. Sommer $^{1}$,$^{4}$}

\affiliation{$^{1}$ Leibniz Institute of Polymer Research Dresden
e.V., 01069 Dresden, Germany}

\affiliation{$^{2}$
 Institute for Condensed Matter
Physics, National Academy of Sciences of Ukraine, UA--79011 Lviv,
Ukraine } \email{zz_usatenko&ph.icmp.lviv.ua} \affiliation{$^{3}$
Marian Smoluchowski Institute of Physics, Jagellonian University,
Department of Statistical Physics, Reymonta 4, Krak\'ow, Poland }
\affiliation{$^{4}$ Institute for Theoretical Physics, Technische
Universit\"at Dresden, 01062 Dresden, Germany}

\date{\today}


\begin{abstract}

The influence of long-range correlated surface and decaying near
surface disorder with quenched defects is studied. We consider a
correlation function for the defects of the form
$\frac{e^{-z/\xi}}{r^{a}}$ , where $a<d-1$ and $z$ being the
coordinate in the direction perpendicular to the surface and $r$
denotes the distance parallel to the surface. We investigate the
process of adsorption of long-flexible polymer chains with excluded
volume interactions on a "marginal" and attractive wall in the
framework of renormalization group field theoretical approach up to
first order of perturbation theory in a double
($\epsilon$,$\delta$)- expansion ($\epsilon=4-d$, $\delta=3-a$) for
the semi-infinite $|\phi|^4$ $O(m,n)$ model with the above mentioned
type of surface and near the surface disorder in the limit $m,n\to
0$. In particular we study two limiting cases. First,
we investigate the scenario where the chain's extension it much larger then $\xi$.
Second, we consider the case where the chain's extension is of the
order of $\xi$. For both cases we obtained series for bulk and the whole set of surface
critical exponents, characterizing the process of adsorption of
long-flexible polymer chains at the surface. The polymer linear
dimensions parallel and perpendicular to the surface and the
corresponding partition functions as well as the behavior of monomer
density profiles and the fraction of adsorbed monomers at the
surface and in the volume are studied.

\end{abstract}

\maketitle

\section{Introduction}
Adsorption and localization processes of macromolecules are very
important for polymer technology in many areas such as lubrication,
adhesion, surface protection or compact disc production.
Investigation of adsorption phenomena allows to understand
biological processes of polymer-membrane interactions as well as
such processes of biotechnology as DNA micro-arrays and
electrophoresis. Therefore, the statistical properties of the
adsorption of long, flexible macromolecular chains (polymers) at
surfaces in dilute, semi-dilute, and concentrated polymer solutions
have found considerable interest \cite{Eisenriegler,Fleer,jones}.

Generally, a thorough theoretical understanding of the statistical
physics of polymers can be reached using well developed mathematical
tools from field theory. Long flexible polymer chains immersed into
a good solvent are very well described by the model of self-avoiding
walks (SAWs) on a regular lattice \cite{Cloizeaux, Schafer}. As
revealed by de Gennes, the conformations of a SAW exhibit critical
(self-similar) behavior approaching the limit of infinite number of
steps which can be extracted from the $m\to 0$ limit of an $O(m)$
symmetric field theory (polymer-magnet analogy)~\cite{deGennes}.
Using this intimate relation with critical phenomena, polymers in
the bulk phase have been extensively studied in the past, and many
of their properties could be clarified \cite{Schafer}. On the other
hand, the theory of surface critical phenomena in such complex
systems has many facets and is still far from being complete. This
concerns in particular adsorption and localization phenomena at
surfaces and interfaces, as well as the influence of different kinds
of disorder effects which naturally occur in many polymer
environments, including structured surfaces.

As noted already by de Gennes \cite{deGennes} and by Barber et al.
\cite{Barber}, there is a formal analogy of the polymer adsorption
problem to the equivalent problem of critical phenomena in the
semi-infinite $|\phi|^4$ $n$-vector model of a magnet with a free
surface \cite{DD81,D86}. In this case any bulk universality class is
split into several surface universality classes, with new surface
critical exponents \cite{DD81,D86}. It should be mentioned that such
analogy takes place for a real polymer chain, i.e., polymers with
excluded volume interaction. Based on the polymer-magnet analogy,
the problem of adsorption of flexible polymer chains on a surface in
the case of good solvent without disorder was investigated few years
ago by Eisenriegler and co-workers
\cite{EKB82,eisenriegler:83,Eisenriegler}. Besides, the adsorption
transition of linear polymers was subject of a series of recent
works, see~\cite{HG94,SKG99,SGK01,VW98,RDGKS02,ZLB90}.

Real systems usually contain some geometric surface defects and
impurities both on the substrate and within the bulk.
On the other hand, disorder effects on surfaces can become
interesting for separation of polymers (electrophoresis on
chip)~\cite{SW95,SB97}, or very general in the context of
nano-structured surfaces and biological systems such as cell membranes.

It has been shown that quenched short-range correlated bulk disorder
does not alter the polymer bulk critical
exponents~\cite{Harris,Kim}. However, the asymptotic behavior of
polymers in media with {\it long-range correlated disorder} is
characterized by a new set of bulk critical
exponents~\cite{Prudnikov,Blavatska2}.

In our recent work \cite{UC04} we have extended previous studies to
investigate the influence of long-range correlated bulk disorder on
the process of adsorption of long flexible polymer chains from
solution onto a planar surface forming the system boundary. Our
investigations have shown that the system with long-range correlated bulk
disorder is characterized  by a new set of surface critical exponents,
and belongs to a new universality class (``long-range'' fixed point).
Our results show
that a larger range of correlations between
the defects enhances the trapping of the chain between
the attractive surface on one side, and the region occupied by the defects on
the other side. Thus, the fraction of the monomers near the wall can be higher than right
at the wall.

Introducing into the system {\it short-range correlated surface
disorder} only turns out to be irrelevant for surface critical
behavior \cite{DN89,PS97,D98}. On the other hand, long-range
correlated surface disorder with a correlation function given  by $
g(r)\sim 1/r^{a} $, where $r$ denotes the distance in the direction
parallel to the surface, can be relevant for the asymptotic behavior
of polymer adsorption if $a<d-1$; but it turns out to be irrelevant
if $a\geq d-1$, as was suggested in Ref.\cite{DN89}. However, there
remain some points to clarify, such as a detailed analysis of the
surface critical behavior of the systems with long-range correlated
surface disorder and the calculation of surface critical exponents.

\smallskip

\begin{figure}[htp]
    \begin{center}
    \includegraphics[width=10cm]{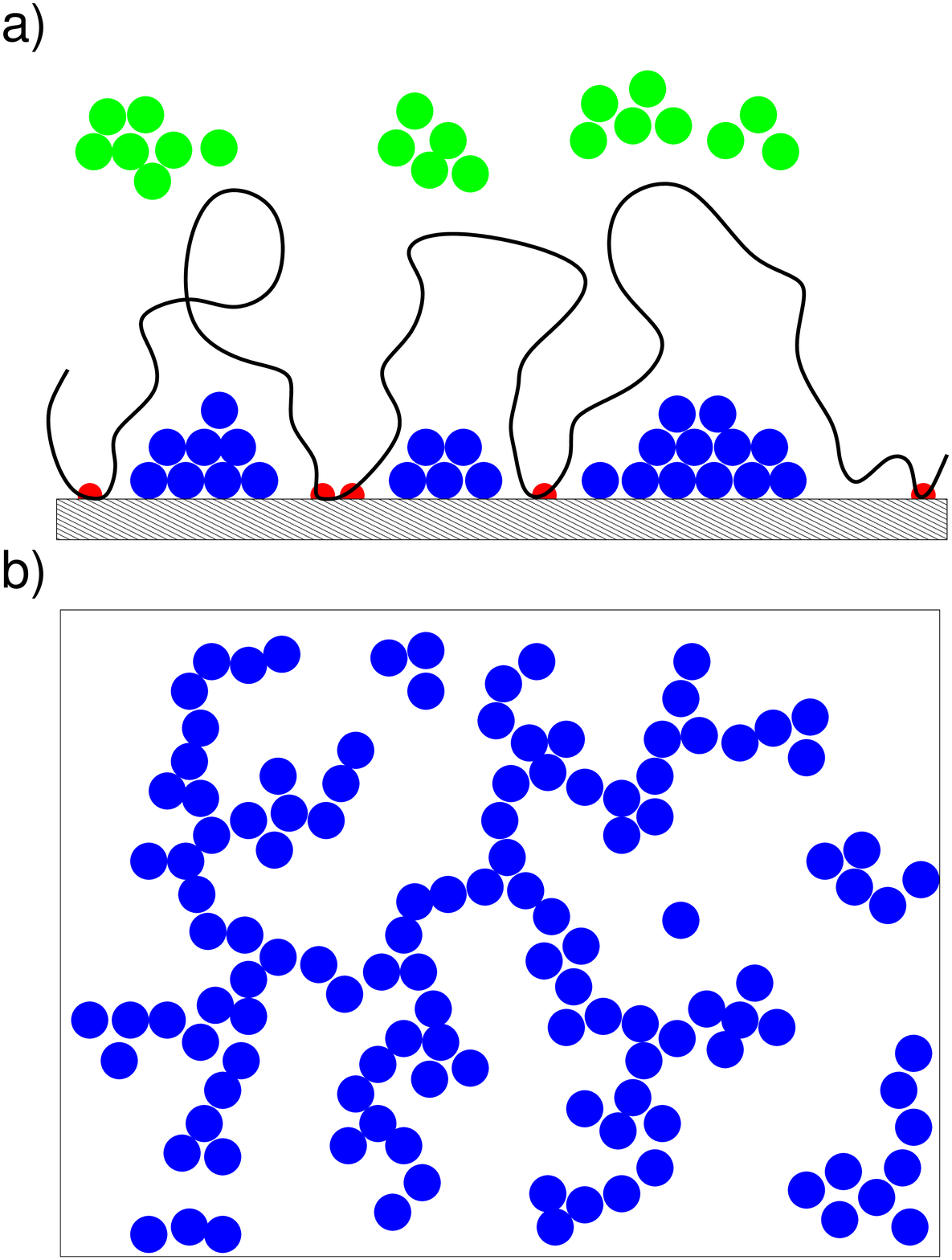}
    \end{center}

\caption{Illustration of the type of disorder considered in this
work. a) side view: Impurities block parts of the surface for
polymer adsorption (blue points) and extent into the bulk (green
points). The chain can only adsorb at sites in between the
impurities (red points). b) Top view: long range correlated disorder
in the direction parallel to the surface is obtained in case of
aggregation processes~\cite{Meakin} or percolation~\cite{Staufer}.
Color online only.} \label{fig:near_surface_disorder}
\end{figure}

In the present work we investigate the influence of quenched
long-range correlated disorder in the direction parallel to the
surface and decaying with a finite correlation length $\xi$ in the
direction perpendicular to the surface. In particular, we consider
such type of disorder with quenched defects obeying by a following
law of correlation :
\begin{equation}\label{g_def}
g(x) = \frac{e^{-z/\xi}}{r^{a}}~~,
\end{equation}
where $z$ denotes the distance between two points in the direction
perpendicular to the surface, and $r$ denotes the distance between
two points in the direction parallel to the surface. Adsorption of
flexible chains is considered on impenetrable surfaces, with $z\ge
0$, in the "marginal" and attractive region of polymer surface
interactions. The model is sketched in
Fig.\ref{fig:near_surface_disorder}. Impurities block surface sites
for polymer adsorption as well as bulk sites. Long range correlated
disorder in the direction parallel to the surface can occur due to
agglomeration or percolation effects of the impurities. The
correlation disappears over a typical length scale of $\xi$ in the
direction perpendicular to the surface. We hasten to note that the
correlation function given in Eq.(\ref{g_def}) implies also a
layer-like structure of the impurities in the bulk.

\smallskip
In order to investigate the critical behaviour of the semi-infinite
$|\phi|^4$ $O(m,n)$ model with above mentioned correlated disorder
in the limit $m,n\to 0$ we apply the renormalization group (RG)
field theoretical approach and choose the massive field theory
scheme with renormalization at non-zero mass and zero external
momenta \cite{Parisi88}. For the quantative analysis of the first
order results we apply the double ($\epsilon$,$\delta$)- expansion.
But it is necessary to add that in our case we use $\epsilon=4-d$,
$\delta=3-a$ expansion in contrast to $\epsilon=4-d$, $\delta=4-a$
expansion proposed by Weinrib and Halperin~\cite{WH83}. The choice
of expansion in $\delta$ is connected in natural way with the value
of $a$ that in our case of surface and near the surface disorder is
$a<d-1$.

The analysis of the obtained series for the surface
critical exponents, characterizing the process of adsorption of
long-flexible polymer chains at the surface is performed. The
polymer linear dimensions parallel and perpendicular to the surface
and the corresponding partition functions as well as the scaling
behavior of monomer density profiles and the fraction of adsorbed
monomers at the surface and close to the surface are studied.

\section{the model}

In this work we investigate the adsorption
phenomena of long, flexible linear polymer chains at a "marginal" and
attractive surface in sufficiently diluted polymeric solution  so that inter-chain interactions
in the bulk and on the surface can be neglected.
In this case the full information about the process of adsorption of
polymer chains at the surface is obtained by
considering the configurations of a single chain.

Let us consider a polymer solution in contact with a
 solid substrate where the monomers in contact with the
 surface gain energy (attractive surface).
Such is usually realized by Van-der-Waals interactions between the
monomers and the substrate. In polarizable polymer-surface systems
also stronger interactions such as hydrogen-bonds can occur. Here,
we will consider the case of monomer-surface interactions which
shall be of the order of $kT$ ($k$ Boltzmann constant, $T$ absolute
temperature). This is usually referred to as weak or reversible
adsorption. By contrast, interaction energies much larger than $kT$
lead to quasi-irreversible adsorption processes which require
non-equilibrium models. Also for weak adsorption at sufficiently low
temperatures, $T<T_a$, an adsorbed state of the chains is caused by
the dominance of surface interactions over the conformational
entropy of free chains, where a finite fraction of the monomers is
localized at the system boundary. This leads to an interesting
transition phenomenon where the polymer chain conformations changes
from an isotropic state to a highly anisotropic "pancake-like"
state.

 The deviation from the adsorption threshold introduced above,
 $c\propto(T-T_a)/T_a$, changes sign at the transition between the adsorbed
 ($c<0$) and the non-adsorbed state ($c>0$) and it plays a role of a
 second critical parameter. The value $1/N$ , where $N$ is a number of monomers,
 plays a role of the primary critical parameter  analogous to the reduced  critical temperature in
 magnetic systems. Thus, the adsorption threshold for infinite
 polymer chains, where $1/N\to 0$ and $c=c_{0}^{sp}\to 0$ is a multicritical
 phenomenon. For $(T>T_a)$ the so-called ordinary transition corresponds
 to a "repulsive surface" and
the limit $1/N \to 0$ leads to usual bulk behavior. The case $T<T_a$
corresponds to the "attractive surface" at which a surface state
occurs. Both regions are joint by the multicritical point which
corresponds to a "marginal" surface state at $T=T_a$. The limit $1/N
\to 0$ at $c = 0$ is referred to as the {\it special transition}. As it
is known \cite{DD81,DSh98}, for each of these regions in the
parameter space the knowledge of one independent surface critical
exponent gives access to the whole set of the other surface critical
exponents via surface scaling relations and the bulk critical
exponents $\nu$ and $\eta$. An example being the critical exponent
$\eta_{\parallel}$ which is related to chain correlations in
directions parallel to the surface.

Of particular importance for polymer adsorption phenomena is the
so-called crossover exponent $\Phi$. The knowledge of $\Phi$ allows
to describe the crossover behavior between the special and ordinary
transitions ($c\ne 0$). The exponent $\Phi$ is related to the length
scale $\xi_c$~\cite{eisenriegler:83,Eisenriegler} given by
\be
\label{xic} \xi_c \sim |c|^{-\nu/\Phi}~~,
\ee
 associated with the parameter $c$.  In the polymer problem the length scale
 $\xi_c$ can be interpreted as the distance from the surface up to
 which the properties of polymer chains depend on the value of $c$, not
 only on its sign.  In the case of adsorption $\xi_c$ defines the extension
of the chain in the direction perpendicular to the surface (blob size).
In the bulk the relevant length scale is the average
 end-to-end distance
 \be
 \label{eq:xi_R}
 \xi_R=\sqrt{<R^2>}\sim N^{\nu}~~.
 \ee
 Additionally, there is the
 microscopic length $l$ -- the statistical segment length which
 denotes the limit of validity of the corresponding coarse grained model.
 Near the multicritical point the only relevant length scales are
 $\xi_R\to\infty$ and $\xi_c\to\infty$. Correspondingly, the properties of the
 system depend on the ratio $\xi_R/\xi_c$. The characteristic ratio is
$(\xi_R/\xi_c)^{\Phi/\nu} \sim |c|N^{\Phi}$, where $cN^{\Phi}$ is
the scaling variable controling weak adsorption of polymers\cite{EKB82}.
On the other hand,
$\Phi$ can be related to the number of monomers in contact to the
surface, $N_1$, at $T_a$ according to $N_1 \sim N^\Phi$, as well as
with the density profile within the chain at $T_a$. There is some
dispute about the correct value of
$\Phi$~\cite{deGennes81,EKB82,deGennesP83,DSh98} for homopolymer
chains at pure surfaces. In a recent work, we have shown that more
reliable results for critical exponents characterizing the process
of adsorption of long-flexible polymer chains can be obtained in the
frames of massive field-theory approach directly at three dimensions
\cite{Usat06}.

\smallskip

 The description of the surface critical behavior
of long-flexible polymer chains near the wall can be formulated in
terms of the effective Landau-Ginzburg-Wilson (LGW) Hamiltonian of
the semi-infinite $m$-vector model \cite{D86,D97}

\be
H = \int_{V} d^{d}x [\frac{1}{2} \mid \nabla\vec{\phi} \mid ^{2}
+ \frac{1}{2} \bar{\mu_{0}}^{2}\mid \vec{\phi} \mid^{2}
+\frac{1}{4!} v_{0} (\vec{\phi}^2)^{2}]+\frac{c_{0}}{2}
\int_{\partial V}d^{d-1}r \vec{\phi}^{2}({\bf{r}},z=0)~~,\label{1}
\ee

where $\vec{\phi}(x)$ is an $m$-vector field with the components
$\phi_{i}(x)$, $i=1,...,m$. It should be mentioned that the
$d$-dimensional spatial integration is extended over a half-space
$V=I\!\!R^d_+\equiv\{{\bf x}{=}({\bf r},z)\in I\!\!R^d\mid {\bf
r}\in I\!\!R^{d-1}, z\ge 0\}$ bounded by a plane free surface
$\partial V$ at $z=0$. Here $\bar{\mu_{0}}^2$ is the "mass", $v_0$ denotes the
coupling constant of the model (excluded volume for polymer chains) and
$c_{0}$ describes the surface-enhancement of the interactions.

Inhomogeneities or defects in the system cause local deviations from
the average value of the transition temperature. This has been shown
in the experiments on Gd~\cite{RE86,WAGSC85,RR87}.
According to the
consideration above, one of the possibilities to introduce disorder
effects into the model is to assume that the parameter
$\bar{\mu_{0}}$ incorporates {\it local random temperature fluctuations}
$\delta\tau(x)$ via $\bar{\mu_{0}}=\mu_{0}+\delta\tau(x)$. The value
$\delta\tau(\bf{x})$ represents the quenched random-temperature
disorder, with $<\delta\tau({\bf{x}})>=0$ and \be
\frac{1}{8}<\delta\tau({\bf{x}})\delta\tau({\bf{x}}')>=g(|
{\bf{x}-\bf{x}'}|), \label{2} \ee
 where angular brackets $<...>$ denote
configurational averaging over quenched disorder. By analogy  with
the isotropic pair correlation function as was introduced by Weinrib
and Halperin \cite{WH83}, we choose the pair correlation function
with anisotropy of disorder in  the direction parallel and
perpendicular to the surface in the form Eq.(\ref{g_def}). The
presence of the surface restricts translational invariance to
translations parallel to the surface. Thus, we introduce the
Fourier-transform ${\tilde{g}}({\bf q},z)$ of $g({\bf x})$ only in
the direction parallel to the surface. Using Eq.(\ref{g_def}), we
obtain

\be {\tilde g}({\bf q},z) \sim u_{0}+w_{0} \mid
{\bf{q}}\mid^{a-d+1}e^{-z/\xi}.\label{3} \ee

Applied to Eq.(\ref{2}) this corresponds to the so-called long-range
correlated "random-temperature" surface and decaying near the
surface disorder. In the case of random uncorrelated point-like (or
short-range correlated) surface disorder the site-occupation
correlation function is $g(r) \sim \delta(r)$ and its
Fourier-transform assumes the simple form ${\tilde g}(q) \sim
u_{0}$. As was shown in \cite{DN89}, short-range correlated (or
random uncorrelated point-like) surface and short-range correlated
bulk disorder \cite{Kim} are irrelevant for surface critical
behavior.  In these cases in the limit $m,n\to 0$ the $u_{0}$ and
$v_{0}$ terms have the same symmetry, and we pass to an effective
Hamiltonian with only one coupling constant $V_{0}=v_{0}-u_{0}$. We
keep the notation $v_{0}$ for the coupling $V_{0}$.

By contrast, taken into account
Eq.(\ref{3}), the long-range correlated surface and near the
surface disorder can be relevant for SAW's, because the $w_{0}$ term
is relevant in the renormalization group sense if $a<d-1$. If $a\ge
d-1$ the $w_{0}$ term is irrelevant and one obtains the effective
Hamiltonian of the system with random uncorrelated point-like
disorder.

Employing the replica trick to carry out averages over different
configurations of the quenched disorder, as it was first explicitly
done in the RG calculations by Grinstein and Luther~\cite{GL76}, it
is possible to construct the effective Hamiltonian of the
semi-infinite $|\phi|^4$ $O(m,n)$ model with a long-range correlated
surface and decaying near the surface disorder

\bea
H_{eff} & = & \sum_{\alpha=1}^{n}\int_{V} d^{d}x
[\frac{1}{2} \mid \nabla\vec{\phi}_{\alpha} \mid ^{2} + \frac{1}{2}
\mu_{0}^{2}\vec{\phi}_{\alpha}^{2} +\frac{1}{4!} v_{0}
(\vec{\phi}_{\alpha}^2)^{2}]\nonumber\\
&-&\sum_{\alpha,\beta=1}^{n}\int d^{d}x_{1} d^{d}x_{2}g(\mid
{\bf{r}}_{1}-{\bf{r}}_{2}\mid,z_2-z_1)\vec{\phi}_{\alpha}^{2}
({\bf{r}}_{1},z_{1})\vec{\phi}_{\beta}^{2}({\bf{r}}_{2},z_{2})\nonumber\\
& + & \frac{c_{0}}{2}\sum_{\alpha=1}^{n} \int_{\partial V}d^{d-1}r
\vec{\phi}_{\alpha}^{2}({\bf{r}},z=0)~~.\label{5}
\eea
Here Greek indices denote replicas, and the replica limit $n\to 0$
is implied.  The limit $m,n\to 0$ of this model describes
the adsorption of long-flexible polymer chains
interacting with a solid substrate in the presence of correlated
defects or impurities. The fields $\phi_{i}({\bf r},z)$ satisfy the
Dirichlet boundary condition in the case of ordinary transition (the
case of repulsive wall): $\phi_{i}({\bf r},z)=0$ at $z=0$ and, the
von Neumann boundary condition in the case of special transition
(the case of "marginal" surface at the adsorption threshold):
$\partial_{z}\phi_{i}({\bf r}, z)=0$ at $z=0$ \cite{DD81,DDE83}. The
model defined in Eq.(\ref{5}) is restricted to translations parallel
to the boundary surface, $z=0$.

\section{Surface critical behavior near the adsorption threshold (fixed point $c_{0}=c_{0}^{ads}$)}
\label{SurfaceCritical}
\subsection{Correlation functions and renormalization conditions}

Correlation functions which involves $N'$ fields
$\phi({\bf{x}}_{i})$ at distinct points ${\bf{x}}_{i}(1\leq i \leq
N')$ in the bulk, $M'$ fields $\phi({\bf{r}}_{j},z=0)\equiv
\phi_{s}({\bf{r}}_{j})$ at distinct points on the wall with parallel
coordinates ${\bf{r}}_{j}(1\leq j \leq M')$, and $L$ insertion of
the bulk operator $\frac{1}{2}\phi^{2}({\bf{X}}_{k})$ at points
${\bf{X}}_{k}$ with $1\leq k \leq L$, $L_{1}$ insertions of the
surface operator $\frac{1}{2}\phi_{s}^{2}({\bf{R}}_{l})$ at points
${\bf{R}}_{l}$ with $1\leq l \leq L_{1}$, have the form
\cite{D86,DSh98}

\be G^{(N',M',L,L_1)}(\{{\bf x}_{i}\},\{{\bf
r}_j\},\{{\bf{X}}_{k}\},\{{\bf{R}}_{l}\}) = < \prod_{i=1}^{N'}
\phi({\bf x}_{i})\prod_{j=1}^{M'}\phi_{s}({\bf
r}_{j})\prod_{k=1}^{L}\frac{1}{2}\phi^{2}({\bf{X}}_{k})
\prod_{l=1}^{L_{1}}\frac{1}{2}\phi^{2}_{s}({\bf{R}}_{l})>~~.
\label{10} \ee

Here, $<...>$ denotes averaging with the Boltzmann factor, where the
Hamiltonian is given in Eq.(\ref{5}). The full free propagator of a
Gaussian chain in {\it semi-infinite space} in the mixed ${\bf{p}},z$
representation is given by \cite{D86} \be G_{0}({{\bf{p}}},z',z) =
\frac{1}{2\kappa_{0}} \left[ e^{-\kappa_{0}|z'-z|} -
\frac{c_{0}-\kappa_{0}}{c_{0}+\kappa_{0}} e^{-\kappa_{0}(z'+z)}
\right],\label{11} \ee
 where
$\kappa_{0}=\sqrt{p^{2}+\mu_{0}^{2}} $ and $\bf p$ denotes the
Fourier transform for the Cartesian components parallel to the
surface.

Taking into account that surface fields $\phi_{s}({\bf{r}})$ and
surface operators $\frac{1}{2}\phi_{s}^{2}(\bf{R})$ scale with
scaling dimensions that are different from those of their bulk
analogs $\phi(\bf{x})$ and $\frac{1}{2}\phi^{2}(\bf{X})$ (see
\cite{DSh98}), the renormalized correlation function involving $N'$
bulk fields and $M'$ surface fields and $L$ bulk operators, $L_{1}$
surface operators  can be written as
\be
 G_{R}^{(N',M',L,L_{1})} ({\bf{p}} ;
\mu,v,w,c)=Z_{\phi}^{-(N'+M')/2} Z_{1}^{-M'/2} Z_{\phi^2}^{L}
Z_{\phi_{s}^2}^{L_{1}} G^{(N',M',L,L_{1})} ({\bf{p}} ;
\mu_{0},v_{0},w_{0},c_{0}),\label{12}
\ee
where $Z_{\phi}$, $Z_{1}$ and $Z_{\phi^2}$, $Z_{\phi_{s}^2}$ are
correspondent UV-finite ($d<4$) renormalization factors. The typical
bulk and surface short-distance singularities of the correlation
functions $G^{(N',M')}$ are removed via mass shift
$\mu_{0}^{2}=\mu^2+\delta \mu^2$ and surface enhancement shift
$c_{0}=c+\delta c$, respectively (see Ref. \cite{DSh98}).

The renormalized mass $\mu$, coupling constants $v$, $w$, and the
renormalization factor $Z_{\phi}$, $Z_{\phi^{2}}$ are fixed formally
via the standard normalization conditions of the infinite-volume
theory \cite{Brezin76,GL76,Par80,PV00}
\bea &&{\Gamma^{(2)}}_{b,R}(q,v,w,\mu)\mid_{q=0}=\mu^2,\nonumber\\
&&{\Gamma^{(4)}}_{v,R}(\{q_{i}\},v,w,\mu)\mid_{\{q_{i}\}=0}=\mu^{\epsilon} v,\nonumber\\
&&{\Gamma^{(4)}}_{w,R}(\{q_{i}\},v,w,\mu)\mid_{\{q_{i}\}=0}=\mu^{\delta}
w,\nonumber\\
&&\frac{\partial}{\partial
q^2}{\Gamma^{(2)}}_{b,R}(q,\mu,v,w)=1,\nonumber\\
&& {\Gamma^{(2,1)}}_{b,R}(\{{\bf q}\},\{{\bf Q}\},\mu,
v,w,)\mid_{\{{{\bf q}={\bf Q}=0}\}}=1,
 \label{br} \eea
with $\epsilon = 4-d$ and $\delta=3-a$. The renormalized vertex
function is given by \be
\Gamma^{(N^{'},L)}_{b,R}(\{q_{i},Q_{i}\},\mu, v,
w)=[Z_{\phi}]^{N^{'}/2}[Z_{\phi^2}]^{L}
\Gamma^{(N^{'},L)}_{b}(\{q_{i},Q_{i}\},\mu_{0}, v_{0},w_{0}). \ee
Thus, taking into account the normalization conditions,
Eq.(\ref{br}), we can remove the typical  bulk short-distance
singularities of the correlation function after performing the mass
renormalization \be
\mu_{0}^{2}=(Z_{\phi})^{-1}\mu^2-\frac{v_{0}}{3}J_{1}(\mu)+\frac{w_{0}}{3}J_{2}(\mu),\label{mr}
\ee with \bea J_{1}(\mu)&=&\frac{1}{(2\pi)^{d-1}}\int
\frac{d^{d-1}q}{2\kappa_{q}},\nonumber\\
J_{2}(\mu)&=& \frac{1}{(2\pi)^{d-1}}\int d^{d-1}q
\frac{|q|^{a-d+1}}{2\kappa_{q}(\frac{1}{\xi}+\kappa_{q})},\label{mr1}
\eea where $\kappa_{q}=\sqrt{q^2+\mu^2}$.
 In order to remove
short-distance singularities of the correlation function
$G^{(N^{'},M^{'})}$ located in the vicinity of the surface, the
surface-enhancement shift $c_{0}=c+\delta c$ is required. The
required surface normalization conditions are (see \cite{DSh98})
\be
G_{R}^{(0,2)}(0;\mu,v,w,c) = \frac{1}{\mu+c}\label{10a}
\ee and
\be
\left.\frac{\partial G_{R}^{(0,2)} (p;\mu,v,w,c)}{\partial
p^{2}} \right|_{p=0} = - \frac{1}{2\mu(\mu+c)^{2}}, \label{11a} \ee
\be
\left.G_{R}^{(0,2,0,1)}({\bf{p}};\mu,v,w,c)\right|_{{\bf{p}}=0}=\frac{1}{(\mu+c)^2}\label{18a}
\ee

Equation (\ref{10a}) defines the surface-enhancement shift $\delta
c$ and shows that the surface susceptibility diverge at $\mu=c=0$.
This point corresponds to the multicritical point
$(\mu_{0c}^{2},c_{0}^{ads})$ at which adsorption threshold takes
place. The normalization condition of Eq. (\ref{11}) and the
expression for the renormalized correlation function of Eq.
(\ref{12}), allow to find the renormalization factor $Z_{\parallel}
= Z_{1} Z_{\phi}$ from the relation

\be
Z_{\parallel}(v,w)^{-1} = \left. 2\mu \frac{\partial}{\partial
p^{2}}[G^{(0,2)} (p)]^{-1}\right|_ {p^2=0} = \lim_{p\to 0}\frac{\mu}
{p}\frac{\partial}{\partial p} [G^{(0,2)}(p)]^{-1}.\label{18}
\ee

The Eq.(\ref{18a}) allows to obtain the renormalization factor
$Z_{\phi_{s}^2}$ from
\begin{equation}
[Z_{\phi_{s}^2}(v,w)]^{-1}=\left. Z_{\parallel}\frac{\partial
[G^{(0,2)}(0;\mu_{0},v_{0},w_{0},c_{0})]^{-1}}{\partial
c_{0}}\right|_{c_{0}=c_{0}(c,\mu,v,w)},\label{18b}
\end{equation}
where relation $
G^{(0,2;0,1)}(0;\mu_{0},v_{0},w_{0},c_{0})=-(\frac{\partial}{\partial
c_{0}})G^{(0,2)}(0;\mu_{0},v_{0},w_{0},c_{0})$ have been taken into
account.

\subsection{Analysis of Callan-Symanzik equations}
Asymptotically close to the critical point
$(\mu_{0c}^{2},c_{0}^{ads})$ the renormalized correlation functions
$G^{(0,2)}$ satisfy the homogeneous Callan-Symanzik (CS) equations
\cite{Z89,Parisi88,ID89} with the corresponding renormalization group (RG) functions.
 The first part of these RG functions are $\beta$-functions
$\beta_{v}(v,w)=\left.\mu\frac{\partial}{\partial \mu}\right|_{0}
v,\quad\beta_{w}(v,w)=\left.\mu\frac{\partial}{\partial
\mu}\right|_{0} w $, and usual bulk exponent $\eta=\left.\mu
\frac{\partial}{\partial{\mu}}\right|_{0}\ln{Z_{\phi}}$. The second
part of these RG functions is a surface-related function
\be
\eta_{1}^{sp}(v,w)=\left.\mu\!\frac{\partial}{\partial
\mu}\right|_{0} \ln\!{Z_{1}}(v,w).\label{19aa}
\ee
In the case of investigation the crossover behavior from the adsorbed to the
non-adsorbed states (see \cite{DSh98,UH03,UC04}) the additional
surface related term arises $-[1+\eta_{\bar{c}}(v,w)]
\bar{c}\frac{\partial}{\partial \bar{c}}$, with crossover-related
function
\be \eta_{\bar{c}}(v,w)=\left. \mu\frac{\partial}{\partial
\mu}\right|_{0} \ln Z_{\phi_{s}^2}(v,w), \label{19b}
\ee
where
$\left.\right|_{0}$ indicates that the derivatives are taken at
fixed  cutoff $\Lambda$, fixed bare coupling constants and surface
enhancement constant.

The simple scaling dimensional analysis of $G_{R}^{(0,2)}$ and of the
mass dependence of the $Z$ factors, allows to express the surface
correlation exponent $\eta_{\parallel}^{sp}$ which characterizes the
critical point correlations parallel to the surface as \be
\eta_{\parallel}^{sp}=\eta_{1}^{sp}+\eta. \label{20} \ee
Taking into account Eqs.~(\ref{18}), (\ref{19aa}) and (\ref{20}), the
surface correlation exponent $\eta_{\parallel}^{sp}$ is presented
via the following expression
\begin{eqnarray}
\eta_{\parallel}^{sp}&=&\left.\mu\!\frac{\partial}{\partial
\mu}\right|_{0}\ln\!{Z_{\parallel}}\nonumber\\
&=&\beta_v(v,w)\frac{\partial\ln Z_{\parallel}(v,w)}{\partial v}+
\left.\beta_w(v,w)\frac{\partial\ln Z_{\parallel}(v,w)}{\partial
w}\,\right|_{F\!P}.\label{21}
\end{eqnarray}
Here FP is a notation of the corresponding fixed point. It should be
mentioned that in the current case the nontrivial long-range (LR)
fixed point is present, which becomes stable when some amount of
long-range correlated surface and near the surface disorder with
$a<d-1$ is introduced into the system.

Taking into account (\ref{19b}), the crossover-related function
$\eta_{\bar{c}}(v,w)$ can be written as \be
\eta_{\bar{c}}(v,w)=\left.\beta_{v}(v,w)\frac{\partial \ln
Z_{\phi_{s}^2}(v,w)}{\partial v}+ \beta_{w}(v,w)\frac{\partial \ln
Z_{\phi_{s}^2}(v,w)}{\partial w}\right|_{F\!P}.\label{27} \ee The
asymptotic scaling critical behavior of the correlation functions
near the multicritical point can be obtained through a detailed
analysis of the CS equations, as was proposed in Ref.
\cite{Z89,BB81} and employed in the case of the semi-infinite
systems in \cite{CR97,DSh98,Sh97, UH03,UC04}.

Using the above mentioned scheme, the asymptotic scaling form of the
surface correlation functions of the long-flexible polymer chains
with one end fixed at the surface and the other end is located
somewhere in the layer $z$ above the surface can be written as \be
G^{\lambda}(z, c_{0})\sim z^{1-\eta^{sp}_{\perp}}G_{\lambda}(\tau
z^{1/\nu}, \tau^{-\Phi} \Delta c_{0})~~.\label{29a} \ee Similarly,
for chains with  both ends fixed on the surface at distance $r$, or
for chains with only one end fixed on the surface and $r_{A}=r_{B}$,
we can write
\begin{equation}
G^{\parallel,\perp}(x;c_{0})\sim
x^{-(d-2+\eta_{\parallel,\perp}^{sp})}G_{\parallel,\perp}(\tau
x^{1/\nu};\tau^{-\Phi}\Delta c_{0}), \label{29b}
\end{equation}
where $\eta_{\perp}^{sp}=\frac{\eta+\eta_{\parallel}^{sp}}{2}$ is
the surface critical exponent which characterizes the critical point
correlations perpendicular to the surface; $\Phi=\nu
(1+\eta_{\bar{c}}(v^{*},w^{*})) $ is the surface crossover critical
exponent \cite{DSh98,UH03,UC04}, which characterizes the measure of
deviation from the multicritical point and $x$ denotes $r$ or $z$ in
$G^{\parallel}$ or $G^{\perp}$, which correspond to $G^{(0,2)}$ and
$G^{(1,1)}$ functions, respectively.

\section{One-loop approximation results}
In general, there are two
possibilities to investigate the critical behavior of the model. In
the first scheme one considers correspondent polynomials for
$\beta$-functions and renormalization factors as functions of
renormalized coupling constants $v,w$ for fixed $d,a$.  Then one
searches for stable solutions of the fixed point equations. The corresponding
one-loop equations in this case do not have any stable accessible
fixed points for $d<4$. In order to obtain reasonable results within
this scheme the knowledge of the second order of perturbation theory
is required. There exists a second scheme to perform the
quantitative analysis of the first order results which implies a
double expansion in $\epsilon=4-d$ and $\delta=3-a$ (with $a<d-1$)
by analogy as was proposed by Weinrib and Halperin \cite{WH83}. But,
it should be mentioned that in our case of long-range correlated
surface and decaying near the surface disorder the upper critical
dimension for the correlation parameter of disorder $a$ is $d_{a}=3$
in contrast to the case of system with long-range correlated bulk
disorder, where $d_{a}=4$. We we use the second approach in our
investigations of adsorption of long-flexible polymer chains with
excluded volume interactions on the surface with long-range
correlated surface and decaying near the surface disorder.

After
performing the integration of the corresponding Feynman diagrams,
for the bulk renormalization factors $Z_{\phi}$ and $Z_{\phi^2}$ at
the first order of perturbation theory, we obtain \bea
Z_{\phi}(\bar{v},\bar{w})&=&1+\frac{\bar{w}}{3}I_{4},\nonumber\\
Z_{\phi^2}(\bar{v},\bar{w})&=&
1+\frac{\bar{v}}{3}+\frac{\bar{w}}{3}(I_{5}-I_{4})\label{brf},
 \eea
where the following definitions for the correspondent integrals were
introduced \bea
I_{4}&=&-(I_{1}\xi)^{-1}\frac{1}{(2\pi)^{d}}\frac{\partial}{\partial
k_{1}^2}\int d^{d}k\frac{|q|^{a-d+1}}{(k_{1}+k)^2+ 1}
\int_{0}^{\infty}dz e^{ifz}e^{-|z|/\xi},\nonumber\\
I_{5}&=&-(I_{1}\xi)^{-1}\mu^{\delta}\frac{1}{(2\pi)^{d}}\frac{\partial}{\partial
\mu^2}\int d^{d}k\frac{|q|^{a-d+1}}{(k_{1}+k)^2+
\mu^2}\int_{0}^{\infty}dz e^{ifz}e^{-|z|/\xi},\label{i4i5}\eea
with
$$ I_{1}=\frac{1}{(2\pi)^{d}}\int \frac{d^{d}k}{(k^2+1)^{2}}=\pi^{-d/2}2^{-d}\Gamma(2-\frac{d}{2})$$
and ${\bf{k}}=({\bf{q}},f)$, where ${\bf{q}}$ is $d-1$ dimensional
vector of momenta. The rescaled renormalized coupling constants
$\bar{v},\bar{w}$ are introduced as follows
 $\bar{v}=v I_{1}$, $\bar{w}=w I_{1}$. Besides, the vertex
 renormalization of the bare parameters in the present case of calculation at first order according to Eq.(\ref{br}) are $v_{0}=v
 \mu^{\epsilon}$ and $-w_{0}\xi=w \mu^{\delta}$ with $\epsilon=4-d$ and
$\delta=3-a$, respectively. It should be mentioned, that in the
present case of anisotropy of disorder in the direction parallel and
perpendicular to the surface we distinguished integration in
direction parallel and perpendicular to the surface. By the
complexity of the initial form of the correlation function
characterizing disorder (see Eq.(\ref{g_def})), the contributions
from parallel momenta ${\bf{q}}$ and perpendicular momenta $f$ are
interconnected which each other. The correspondent $\beta$-functions
has a form \bea \beta_{\bar{v}}(\bar{v},\bar{w})&=&-\epsilon \bar{v}
(1-\frac{4}{3}\bar{v})+2\delta\bar{v}\bar{w}
(I_{2}-\frac{I_{4}}{3})+\frac{2}{3}{\bar{w}}^2 I_{3}(2\delta-\epsilon),\nonumber\\
\beta_{\bar{w}}(\bar{v},\bar{w})&=&-\delta
\bar{w}+\frac{2}{3}\bar{v}\bar{w}\epsilon+\frac{2}{3}\delta
{\bar{w}}^2(I_{2}-I_{4}),\label{sd3} \eea

where we have introduced the following definitions
\bea I_{2}&=&
(I_{1}\xi)^{-1}\frac{1}{(2\pi)^{d}}\int
d^{d}k\frac{|q|^{a-d+1}}{((k_{1}+k)^2+ 1)^{2}} \int_{0}^{\infty}dz
e^{ifz}e^{-|z|/\xi},\nonumber\\
I_{3}&=&(I_{1}\xi)^{-1}\frac{1}{(2\pi)^{d}}\int
d^{d}k\frac{|q|^{2(a-d+1)}}{((k_{1}+k)^2+ 1)^{2}}
\int_{0}^{\infty}dz e^{ifz}e^{-2|z|/\xi}.\label{i2i3} \eea

Taking into account (\ref{brf}) and correspondent order of
$\beta$-functions we obtain one-loop order results for the bulk
critical exponents $\nu$ and $\eta$
 \bea
\nu&=&\frac{1}{2}+\frac{\epsilon {\bar{v}}}{12}+ \frac{\delta
{\bar{w}}}{12}(I_{5}-I_{4})
 ,\nonumber\\
\eta&=& -\frac{{\bar{w}}\delta}{3}I_{4}.\label{nu} \eea

 In our calculations we distinguish two cases: near the surface disorder
$\xi<<\xi_{R}$ and the case of extended disorder on the distance of
the correlation length $\xi_{R}$, i.e. $\xi\sim\xi_{R}$.

\subsection{The case of near the surface disorder $\xi<<\xi_{R}$}
 First of all discuss the situation of
near the surface disorder, i.e. the case $\xi<<\xi_{R}$.

After performing double $(\epsilon,\delta)$ - expansion of the above
mentioned integrals Eq.(\ref{i4i5}) and Eq.(\ref{i2i3}), we obtain
\be I_{4}= \sim \frac{(\delta-\epsilon)\epsilon} {2\delta^{2}
\pi^{\frac{d}{2}+1}},\quad\quad\quad I_{5}= \sim
\frac{(1-\delta)\epsilon}{\delta},\label{sd2} \ee and $$ I_{2}\sim
\frac{\epsilon}{\delta},\quad\quad\quad I_{3}\sim
\frac{\epsilon}{2(2\delta-\epsilon)}.$$

The general forms of integrals $I_{4}$ and $I_{5}$ for the above
mentioned case $\xi<<\xi_{R}$ are presented in Appendix 1.
 The integration of the correspondent Feynman integrals in the
renormalized two-point correlation function of two surface fields
$G^{(0,2)}$ at the first order of the perturbation theory gives for
the surface renormalization factors $Z_{\parallel}$ and
$Z_{\phi_{s}^{2}}$  the following results
 \be
Z_{\parallel}=1+\frac{\bar{v}}{3(1+\epsilon)}+\frac{2\bar{w}
I_{6}}{3}(1-\frac{I_{4}}{4I_{6}}),\label{Zpar}\ee \bea
Z_{\phi_{s}^{2}}&=&1-\frac{\bar{v}}{3(1+\epsilon)}(1-2^{\frac{1+\epsilon}{2}}{}_{2}F_{1}
[\frac{3-\epsilon}{2},\frac{\epsilon+1}{2},\frac{3+\epsilon}{2},\frac{1}{2}])\nonumber\\
&-&\frac{2\bar{w}
I_{6}}{3}(1-\frac{I_{4}}{4I_{6}}-2^{\frac{1+\delta}{2}}
{}_{2}F_{1}[\frac{3-\delta}{2},\frac{\delta+1}{2},\frac{3+\delta}{2},\frac{1}{2}]),
\label{sd1} \eea

where integrals $I_{6}=I_{5}/(1+\delta)$ and the function
$_{2}F_1[...]$ is a standard Hypergeometric function.

 Combining the renormalization factors $Z_{\parallel}$ and $Z_{\phi_{s}^{2}}$ together with
 the corresponding order of  $\beta$-functions Eq.(\ref{sd3})
the surface critical exponents $\eta_{\parallel}$ and
$\eta_{\bar{c}}$ according to (\ref{21}), (\ref{27}) can be obtained
\bea \eta_{\parallel}(\bar{v},\bar{w})&=&-\frac{\epsilon
\bar{v}}{3(1+\epsilon)}-\frac{2}{3}\delta \bar{w} I_{6}(1-\frac{I_4}{4I_{6}}),\nonumber\\
\eta_{\bar{c}}(\bar{v},\bar{w})&=&\frac{\epsilon
\bar{v}}{3(1+\epsilon)}(1-2^{\frac{1+\epsilon}{2}}{}_{2}F_{1}
[\frac{3-\epsilon}{2},\frac{\epsilon+1}{2},\frac{3+\epsilon}{2},\frac{1}{2}])+\nonumber\\
&+&\frac{2}{3}\delta \bar{w}
I_{6}(1-\frac{I_{4}}{4I_{6}}-2^{\frac{1+\delta}{2}}{}_{2}F_{1}
[\frac{3-\delta}{2},\frac{\delta+1}{2},\frac{3+\delta}{2},\frac{1}{2}]).
\label{sd4} \eea

 It should be mentioned that in the case
$\xi=0$ which corresponds to the situation of only surface disorder
from Eq.(\ref{sd3}) follows that pure bulk fixed point becomes
stable.

In the limit $\epsilon\to 0^{+}$ for the exponent functions
$\eta_{\parallel}$ and $\eta_{\bar{c}}$ we obtain \be
{\lim_{\epsilon\to 0^{+}}}\eta_{\parallel}={\lim_{\epsilon\to
0^{+}}}\eta_{c}\sim-\frac{\tilde{v}}{3}-\frac{2}{3}
{\tilde{w}}\frac{(1-\delta)}{1+\delta},\label{sd5} \ee where the
following definitions are introduced: $\tilde{v}=vK_{4}$ and
$\tilde{w}=wK_{4}$ with $K_{4}=1/(8\pi^2)$.
 In the case $\epsilon \to 1$ (i.e. $d=3$) the Eq.(\ref{sd4}) lead to
\bea \eta_{\parallel}&=&-\frac{\bar{v}}{6}-\frac{2}{3}{\bar{w}}\frac{(1-\delta)}{1+\delta}(1-I_{7}),\nonumber\\
\eta_{\bar{c}}&=&\frac{{\bar{v}}}{6}(1-4\ln 2)+\frac{2}{3}
{\bar{w}}\frac{(1-\delta)}{1+\delta}(1-I_{7}-2^{\frac{1+\delta}{2}}{}_{2}F_{1}[\frac{3-\delta}{2},\frac{1+\delta}{2},\frac{3+\delta}{2},\frac{1}{2}]),\label{sd6}
\eea where $I_{7}\sim \frac{(1+\delta)}{4\pi^{5/2}}$. In the special
case $\delta=1$, which corresponds to the case of short-range
correlated (or random uncorrelated point-like) surface disorder,
from (\ref{sd5}) and (\ref{sd6}) we obtain the surface critical
exponents $\eta_{\parallel}$ and $\eta_{\bar{c}}$ of the pure model.
These results confirm predictions that short-range correlated
surface disorder is irrelevant for surface critical behavior and
have not any influence on the process of adsorption of polymer
chains.

The fixed points (${\bar{v}}^{*},{\bar{w}}^{*}$) are given by
solutions of the system of equations:
$\beta_{\bar{v}}({\bar{v}}^{*},{\bar{w}}^{*})=0,\beta_{\bar{w}}({\bar{v}}^{*},{\bar{w}}^{*})=0$.
The stable fixed point is defined as the fixed point where the
stability matrix \be
\begin{pmatrix} \frac{\partial\beta_{\bar{v}}}{\partial{\bar{v}}}&
\frac{\partial\beta_{\bar{v}}}{\partial{\bar{w}}}\\
\frac{\partial\beta_{\bar{w}}}{\partial{\bar{v}}}&\frac{\partial
\beta_{\bar{w}}}{\partial{\bar{w}}}
\end{pmatrix}
\ee hold eigenvalues $\lambda_{\bar{v}}$, $\lambda_{\bar{w}}$ with
positive real parts. In general, there are three accessible fixed
points: the Gaussian (G) fixed point
${\bar{v}}^{*}=0,{\bar{w}}^{*}=0$, the pure (P) SAW fixed point
${\bar{v}}^{*}=3/4, {\bar{w}}^{*}=0$ (here we keep definitions
${\bar{v}}$ and ${\bar{w}}$ for rescaled renormalized coupling
constant introduced above) and one of two LR mixed fixed points with
${\bar{v}}_{i}^{*}\ne 0, {\bar{w}}_{i}^{*}\ne 0$, where $i=1,2$. The
Gaussian fixed point with $\lambda_{1}=-\epsilon$ and
$\lambda_{2}=-\delta$ is never stable for positive $\epsilon$ and
$\delta$. The pure fixed point is stable for systems without
disorder and in the case of positive real parts
$\lambda_{1}=\epsilon$ and $\lambda_{2}=\epsilon/2-\delta$. When we
introduce long-range correlated surface and near the surface
disorder into the system one of the LR mixed fixed points, obtained
in the framework of $(\epsilon,\delta)$- expansion,
${\bar{v}}^{*}=\frac{3}{4}\frac{2\delta^2}{\epsilon(\epsilon-\delta)},
{\bar{w}}^{*}=\frac{3}{2}\frac{\delta(\epsilon-2\delta)}{\epsilon(\epsilon-\delta)}$
becomes stable for $\delta<\epsilon<2\delta$ and $a=3-\delta<d-1$.

The above values of the surface critical exponents Eq.(\ref{sd6})
$\eta_{\parallel}$ and $\eta_{\bar{c}}$ formally should be
calculated at the above mentioned LR mixed fixed point
(${\bar{v}}^{*}\ne 0,{\bar{w}}^{*}\ne 0$).

The other surface critical exponents can be calculated on the basis
of the surface scaling relations (see Appendix) and series for the
bulk critical exponents $\nu$ and $\eta$.

In the special case $\delta=1$ from Eq.(\ref{nu}) and Eq.(\ref{sd4})
taking into account Eq.(\ref{sd2})) we obtain the result for
critical exponent of pure model \cite{Cloizeaux,deGennes}.
Substituting the above mentioned mixed fixed point ${\bar{v}}^{*}\ne
0$ and ${\bar{w}}^{*}\ne 0$ into Eq.(\ref{nu}) the first order
result for critical exponent $\nu$ read:

\be \nu=\frac{1}{2}+\frac{\delta}{8}.\label{nuexp} \ee This result
formally coincides with previous results for the case of long-range
correlated bulk disorder (see (\cite{Blavatska2})), but it should be
mentioned that in our case of long-range correlated surface and
decaying near the surface disorder the correlation parameter of
disorder $a=3-\delta$ has another upper critical dimensions and is
valid for $a<d-1$. The correlation parameter $\delta$ characterizes
long-range correlated disorder in the direction parallel to the
surface and parameter $\xi$ describes disorder in the direction
perpendicular to the surface. The influence of the exponentially
decaying type of disorder in the direction perpendicular to the
surface reduces to the renormalization of the coupling constant
$w_{0}$ via $-w_{0}\xi=w\mu^{\delta}$, as was indicated before. In
accordance with the complexity of the initial form of the
correlation function characterizing disorder (see Eq.(\ref{g_def})),
the contributions from parallel and perpendicular part are
interconnected which each other and can not be split off. The
critical exponent $\nu$ describes the overall swelling of the
polymer coil and increases when the correlation of the disorder is
increased (i.e., $a=3-\delta$ decreased).

\subsection{The case of extended disorder $\xi\sim\xi_{R}$}

After performing double $(\epsilon,\delta)$ - expansion of the above
mentioned integrals Eq.(\ref{i4i5}) and Eq.(\ref{i2i3}), in the case
of extended disorder on the distance of the correlation length
$\xi_{R}$, i.e. $\xi\sim\xi_{R}$ we obtain

$$ I_{2}\sim \frac{\epsilon(1-\delta)}{(1+\delta)},\quad\quad
I_{3}\sim \frac{\epsilon}{3}(1+\frac{4}{3}(\epsilon-2\delta)),$$ and
\be I_{4}\sim\frac{(\delta-\epsilon)\epsilon}{2\delta
\pi^{\frac{d}{2}+1}(1+\delta)},\quad\quad\quad I_{5}
\sim\frac{\epsilon(1-\delta)}{(1+\delta)}.\label{i4kxb}\ee

The general forms of integrals $I_{4}$ and $I_{5}$ for the present
case $\xi\sim\xi_{R}$ are presented in Appendix 2. In the present
case we have the Gaussian (G) fixed point
${\bar{v}}^{*}=0,{\bar{w}}^{*}=0$, the pure (P) SAW fixed point
${\bar{v}}^{*}=3/4, {\bar{w}}^{*}=0$ and two LR mixed fixed points
with ${\bar{v}}_{i}^{*}\ne 0, {\bar{w}}_{i}^{*}\ne 0$, where
$i=1,2$. When we introduce long-range correlated surface and near
the surface disorder Eq.(\ref{g_def}) with $\xi\sim\xi_{R}$ into the
system one of the LR mixed fixed points
${\bar{v}}^{*}=\frac{3}{4}\frac{2\delta}{\epsilon},
{\bar{w}}^{*}=\frac{3}{2}\frac{1}{2\epsilon}\sqrt{\frac{3}{8}}(1-\delta)$
becomes stable for $1<\delta$ and $\epsilon<4\delta$. Fortunately,
in this case $\xi\sim\xi_{R}$ the LR fixed point is stable in the
region where power counting in Eq.(\ref{3}) shows that such type of
disorder is relevant. For the critical exponent $\nu$ in this case
we obtain \be \nu=
\frac{1}{2}+\frac{\delta}{8}(1+\frac{1}{2}\sqrt{\frac{3}{8}}).\label{nu2}\ee
In the present case of $\xi\sim\xi_{R}$ the critical exponent $\nu$
contains additional contribution proportional to $\delta$ in
comparison with the case $\xi<<\xi_{R}$. The dependence $\nu$ from
$\delta$ presented in Table 1. As it is easy to see from this table,
the values of critical exponent $\nu$ increase, when the correlation
of the disorder increase (i.e., $a=3-\delta$ decrease). The
integration of the correspondent Feynman integrals in the
renormalized two-point correlation function of two surface fields
$G^{(0,2)}$ at the first order of the perturbation theory gives for
the surface renormalization factors $Z_{\parallel}$ and
$Z_{\phi_{s}^{2}}$ in the present case of $\xi\sim\xi_{R}$ the
following results
 \be
Z_{\parallel}=1+\frac{\bar{v}}{3(1+\epsilon)}+\frac{2\bar{w}
I^{*}_{6}}{3}(1-\frac{I_{4}}{4I^{*}_{6}}),\label{Zpar1}\ee \bea
Z_{\phi_{s}^{2}}&=&1-\frac{\bar{v}}{3(1+\epsilon)}(1-2^{\frac{1+\epsilon}{2}}{}_{2}F_{1}
[\frac{3-\epsilon}{2},\frac{\epsilon+1}{2},\frac{3+\epsilon}{2},\frac{1}{2}])\nonumber\\
&-&\frac{2}{3}\bar{w}
I_{6}(1-\frac{I_{4}}{4I_{6}}-2^{\frac{1+\delta}{2}} I_{9}),
\label{sd1b} \eea

where we introduced the next definitions of $I^{*}_{6}$, $I_{8}$ and
$I_{9}$ \be
I^{*}_{6}=I_{6}(1-2^{\frac{\delta-1}{2}}3^{\frac{1-\delta}{2}}{}_{2}F_{1}[\frac{3-\delta}{2},\frac{1+\delta}{2},\frac{3+\delta}{2},-\frac{1}{2}]),
\quad\quad I_{8}=\frac{(1+\delta)\Gamma(\delta-1)} {2\Gamma(\delta)},\nonumber\\
\ee \bea
I_{9}&=&{}_{2}F_{1}[\frac{3-\delta}{2},\frac{1+\delta}{2},\frac{3+\delta}{2},\frac{1}{2}]+
\frac{3^{\frac{1-\delta}{2}}}{2}{}_{2}F_{1}[\frac{3-\delta}{2},\frac{1+\delta}{2},\frac{3+\delta}{2},-\frac{1}{2}]\nonumber\\
&+&I_{8}(3^{\frac{1-\delta}{2}}{}_{2}F_{1}[\frac{3-\delta}{2},\frac{\delta-1}{2},\frac{1+\delta}{2},-\frac{1}{2}]
-{}_{2}F_{1}[\frac{3-\delta}{2},\frac{\delta-1}{2},\frac{1+\delta}{2},\frac{1}{2}])
\eea
 Combining the renormalization factors $Z_{\parallel}$ and $Z_{\phi_{s}^{2}}$ together with
 the corresponding order of  $\beta$-functions Eq.(\ref{sd3})
the surface critical exponents $\eta_{\parallel}$ and
$\eta_{\bar{c}}$ according to (\ref{21}), (\ref{27}) can be obtained
\bea \eta_{\parallel}(\bar{v},\bar{w})&=&-\frac{\epsilon
\bar{v}}{3(1+\epsilon)}-\frac{2}{3}\delta \bar{w} I^{*}_{6}(1-\frac{I_4}{4I^{*}_{6}}),\nonumber\\
\eta_{\bar{c}}(\bar{v},\bar{w})&=&\frac{\epsilon
\bar{v}}{3(1+\epsilon)}(1-2^{\frac{1+\epsilon}{2}}{}_{2}F_{1}
[\frac{3-\epsilon}{2},\frac{\epsilon+1}{2},\frac{3+\epsilon}{2},\frac{1}{2}])+\nonumber\\
&+&\frac{2}{3}\delta \bar{w}
I_{6}(1-\frac{I_{4}}{4I_{6}}-2^{\frac{1+\delta}{2}}I_{9}).
\label{sd4b} \eea

In the limit $\epsilon\to 0$ in the present case for the exponent
functions $\eta_{\parallel}$ and $\eta_{c}$ we obtain

\bea \lim_{\epsilon\to
0}\eta_{\parallel}&=&-\frac{\tilde{v}}{3}-\frac{2}{3}\tilde{w}\frac{(1-\delta)}{1+\delta}(1-2^{\frac{\delta-1}{2}}3^{\frac{1-\delta}{2}}
{}_{2}F_{1}[\frac{3-\delta}{2},\frac{1+\delta}{2},\frac{3+\delta}{2},-\frac{1}{2}]-\frac{\delta}{8\pi^3}),\nonumber\\
\lim_{\epsilon\to 0}\eta_{c}&=&-\frac{\tilde{v}}{3}+\frac{2}{3}
\tilde{w}\frac{(1-\delta)}{(1+\delta)}(1-2^{\frac{1+\delta}{2}}I_{9}-\frac{\delta}{8\pi^{3}}),
\label{eta11etac} \eea
 In the limit case $\epsilon\to 1$ (i.e. $d=3$) from
Eq.(\ref{sd4b}) we obtain

\bea \lim_{\epsilon\to
1}\eta_{\parallel}&=&-\frac{\bar{v}}{6}-\frac{2}{3}\bar{w}\frac{(1-\delta)}{(1+\delta)}(1-2^{\frac{\delta-1}{2}}3^{\frac{1-\delta}{2}}
{}_{2}F_{1}[\frac{3-\delta}{2},\frac{1+\delta}{2},\frac{3+\delta}{2},-\frac{1}{2}]-\frac{\delta}{8\pi^2}),\nonumber\\
\lim_{\epsilon\to 1}\eta_{c}&=&\frac{\bar{v}}{6}(1-4\ln
2)+\frac{2}{3}\tilde{w}\frac{(1-\delta)}{(1+\delta)}(1-2^{\frac{1+\delta}{2}}I_{9}-\frac{\delta}{8\pi^{2}}).\label{etablime1}\eea

In the limit case $\delta\to 1$, which corresponds to the situation
of short-range correlated (or random uncorrelated point-like)
surface disorder (where $a=3-\delta$) from Eqs.(\ref{eta11etac}) and
(\ref{etablime1}) we obtain the surface critical exponents of pure
model.
 The results of calculation of surface critical
exponents $\eta_{\parallel}$ and $\eta_{\bar{c}}$ presented in Table
1. As it is easy to see from this table, the values of surface
critical exponent $\eta_{\parallel}$ and $\eta_{\bar{c}}$ decrease,
when the correlation of the disorder increase (i.e., $a=3-\delta$
decrease) in comparison with first order results
$\eta_{\parallel}=-0.13$, $\eta_{\bar{c}}=-0.22$ and one-loop order
results $\eta_{\parallel}=-0.204$, $\eta_{\bar{c}}=-0.362$ for pure
model without disorder (see \cite{DSh98}).

\section{Scaling analysis}\label{Scaling}

The knowledge of the above mentioned surface critical exponents
$\eta_{\parallel}$ and $\Phi$ and bulk critical exponents $\nu$ and
$\eta$ is sufficient for the scaling analysis of
different characteristics near the multicritical point $1/N\to 0$
and $c\to 0$. So, it allows us to investigate behavior of long-flexible polymer chains near
the  marginal and adsorbing  surface. Furthermore,
the crossover from adsorbed to desorbed state can be analyzed.

Let us first consider the mean square end-to-end distance of a chain with one end
attached to the surface and the other one freely fluctuating. In the semi-infinite
system translational invariance is broken, and the parallel $
<R^2_{\parallel}>$ and perpendicular $<R^2_{\perp}>$ parts of the
average end-to-end distance $<R^2>=<R^2_{\perp}+R^2_{\parallel}>$
 should be distinguished. The perpendicular part $<R^2_{\perp}>^{1/2}$
 in the case $c\ge 0$ is proportional to  $N^{\nu}$ and has the
 same value as the asymptotic behavior in the bulk.
 In the adsorbed state, $c<0$, the part $<R^2_{\perp}>^{1/2}$ is independent of $N$
and describes the thickness $\xi_{th}$ of the adsorbed chain: \be
\label{xi} \xi_{th} = <R^2_{\perp}>^{1/2}\sim\xi_c \hskip1cm c<0.
\ee

This thickness diverges for $c=1/N=0$, see Eq.(\ref{xic}). According
to Eq.(\ref{xic}) this thickness is controlled by the crossover
exponent $\Phi$ and thus depends crucial of the type of disorder
(see Table II). This dependence of the thickness of the adsorbed
layer from $c$ for different values of correlation parameter
$\delta=3-a$ is presented on Figure 2.

\begin{figure}[htb]
\begin{center}
\includegraphics[scale=0.8]{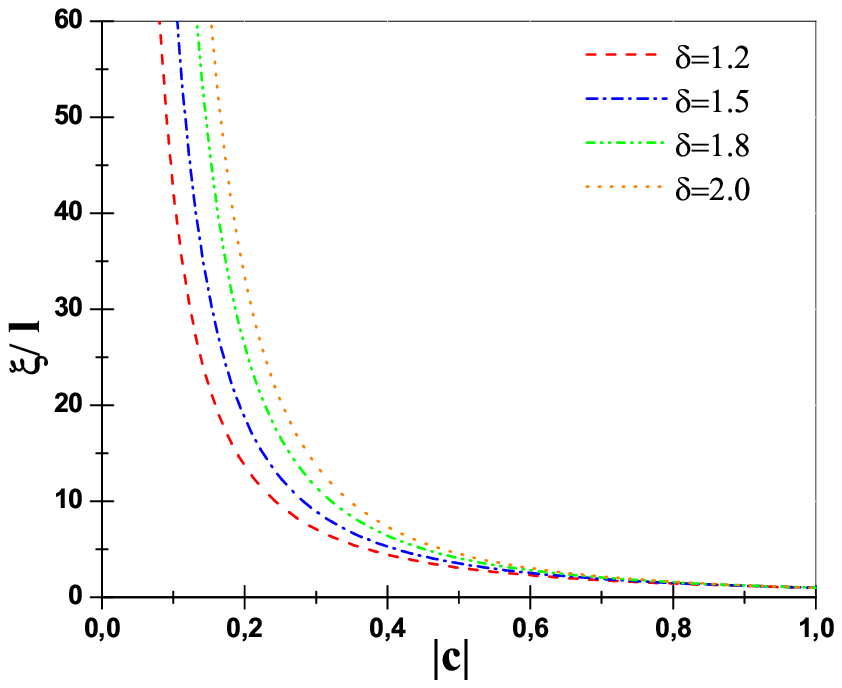}
\end{center}
\caption{The dependence of the thickness of the adsorbed layer on
$c$ for different values of correlation parameter $\delta=3-a$ for
$c < 0$(i.e. below the adsorption threshold). Both quantities are
dimensionless.} \label{fig2}
\end{figure}

 The asymptotic scaling form of $ <R^2_{\parallel}>^{1/2}$ for $c<0$ is
 $<R^2_{\parallel}>^{1/2}\sim
 |c|^{(\nu^{d-1}-\nu)/\Phi}N^{\nu^{d-1}}$, where $\nu^{d-1}$ is the
 correlation exponent in $d-1$ dimensions. For $c\ge 0$ the scaling
 form of $<R^2_{\parallel}>^{1/2}$ is proportional to $N^{\nu}$,
 i.e. it is also the same as in the bulk.

As mentioned in Ref.~\cite{EKB82}, the knowledge of Eqs.(\ref{29a})
and (\ref{29b}) gives access to the short-distance behavior for
$l\ll
 z,r\ll N^{\nu}$ at the adsorption threshold of the corresponding partition functions with one
end fixed and another end free and for the partition function with
two ends at the surface
\begin{equation}
Z^{\lambda}(0,z)\sim z^{a_{\lambda}} N^{b_{\lambda}},\label{Zl}
\end{equation}
\begin{equation}
Z^{\parallel,\perp}(x)\sim x^{a_{\parallel,\perp}}
N^{b_{\parallel,\perp}},\label{Zpp}
\end{equation}
where critical exponents are: $a_{\lambda}=
\eta_{\parallel}-\eta_{\perp}$, $b_{\lambda}=-1+\gamma_{\parallel}$
and  $a_{\parallel,\perp}=1-\eta_{\parallel,\perp}-\Phi/\nu$,
$b_{\parallel,\perp}=-1-\nu (d-1)+\Phi$. Table II represents the
obtained  first order results for the corresponding critical
exponents characterizing the process of adsorption of long-flexible
polymer chains near the adsorption threshold and their dependence
from the correlation parameter $\delta$.
We note that the values for the cross-over exponent are much smaller than
for the case of pure surfaces without any disorder effects. This, however,
is compensated by the large values of Flory exponent.

\begin{table}[htb]
\caption{\label{tab:tab1}Bulk and surface critical exponents
characterizing the process of adsorption of long-flexible polymer
chains at the adsorption threshold $c=c_a$ and in the crossover
region between adsorbed and desorbed states calculated for different
fixed values of the correlation parameter $\delta=3-a$ (in the limit
$\epsilon\to 1$).}
\begin{center}
\begin{tabular}{rrrrrrrrr}
\hline
$ \delta $~&~$ \nu $~&~$ \gamma $~&~$ \eta_{\parallel}$~&~$
\eta_{\perp}$~&~$ \gamma_{1} $~&~$ \gamma_{\parallel} $~&~$ \eta_{c} $~&~$ \Phi $\\
 \hline
 1.1  & 0.680  &  1.364  & -0.271 & -0.142 & 1.430 & 0.815 & -0.492 & 0.434 \\

 1.2  & 0.696  &  1.397  & -0.295 & -0.155 & 1.469 & 0.844 &-0.536 & 0.428 \\

 1.3  & 0.712  &  1.430  & -0.320 & -0.168 & 1.508 & 0.872 & -0.581 & 0.422 \\

 1.4  & 0.729  &  1.463  & -0.345 & -0.180 & 1.547 & 0.901 &-0.626 & 0.416 \\

 1.5  & 0.745  &  1.496  & -0.369 & -0.193 & 1.587 & 0.930 &-0.671 & 0.410 \\

 1.6  & 0.761  &  1.529  & -0.394 & -0.206 & 1.626 & 0.958 &-0.715 & 0.404 \\

 1.7  & 0.778  &  1.562  & -0.418 & -0.219 & 1.665 & 0.987 &-0.760 & 0.398 \\

 1.8  & 0.794  &  1.595  & -0.443 & -0.232 & 1.704 & 1.015 &-0.805 & 0.392 \\

 1.9  & 0.810  &  1.628  & -0.468 & -0.245 & 1.743 & 1.044 &-0.849 & 0.386 \\

 2.0  & 0.827  &  1.661  & -0.492 & -0.258 & 1.782 & 1.073 &-0.894 & 0.380 \\

\end{tabular}
\end{center}
\end{table}

\begin{table}[htb]
\caption{\label{tab:tab2}Critical exponents characterizing the
process of adsorption of long-flexible polymer chains near the
adsorption threshold $c=c_a$ calculated for different fixed values
of the correlation parameter $\delta=3-a$ (in the limit $\epsilon\to
1$).}
\begin{center}
\begin{tabular}{rrrrrrrrrrrrr}
\hline
$ \delta $~&~$ \frac{(1-\Phi)}{\nu}  $~&~$ \frac{\nu}{\Phi}
$~&~$ \frac{(1-\Phi)}{\Phi} $~&~$ \bar{a}$~&~$ \bar{b} $~&~$ a^{'}$~&~$ b^{'} $~&~$ a_{\lambda}$~&~$ b_{\lambda}$~&~$ a_{\parallel} $~&~$ a_{\perp}$~&~$ b_{\parallel,\perp}$ \\
 \hline
 1.1  & 0.833  &  1.567  & 1.306 & 0.167 & 0.864 & -0.098 & 0.430 & -0.129 & -0.185 & 0.633 & 0.504 & -2.793 \\

 1.2  & 0.822  &  1.627  & 1.338 & 0.178 & 0.897 & -0.104 & 0.469 & -0.141 & -0.156 & 0.681 & 0.540 & -2.820 \\

 1.3  & 0.812  &  1.689  & 1.371 & 0.188 & 0.930 & -0.111 & 0.508 & -0.153 & -0.128 & 0.728 & 0.576 & -2.846 \\

 1.4  & 0.802  &  1.753  & 1.406 & 0.198 & 0.963 & -0.116 & 0.547 & -0.164 & -0.099 & 0.774 & 0.610 & -2.873 \\

 1.5  & 0.793  &  1.819  & 1.441 & 0.207 & 0.996 & -0.122 & 0.587 & -0.176 & -0.071 & 0.819 & 0.643 & -2.899 \\

 1.6  & 0.784  &  1.886  & 1.478 & 0.216 & 1.029 & -0.127 & 0.626 & -0.188 & -0.042 & 0.864 & 0.676 & -2.926 \\

 1.7  & 0.775  &  1.956  & 1.515 & 0.225 & 1.062 & -0.132 & 0.665 & -0.199 & -0.013 & 0.907 & 0.708 & -2.953 \\

 1.8  & 0.766  &  2.028  & 1.554 & 0.234 & 1.095 & -0.137 & 0.704 & -0.211 &  0.015 & 0.950 & 0.739 & -2.979 \\

 1.9  & 0.759  &  2.102  & 1.594 & 0.241 & 1.128 & -0.142 & 0.743 & -0.223 &  0.044 & 0.992 & 0.769 & -3.006 \\

 2.0  & 0.751  &  2.178  & 1.635 & 0.249 & 1.162 & -0.147 & 0.782 & -0.234 &  0.073 & 1.033 & 0.799 & -3.033 \\

\end{tabular}
\end{center}
\end{table}

For the fraction of monomers at the surface, $N_1/N$, the asymptotic
behavior is closely related to the crossover
exponent~\cite{EKB82,Eisenriegler}:
 \begin{eqnarray}
\label{N1/N} N_{1}/N\sim \left\{
 \begin{array}{lll}
|c|^{(1-\Phi)/\Phi} &\;\; \mbox{if}  & \;\;\; \mbox{$c<0$}\\ \label{c_le_0}
N^{\Phi-1} &\;\; \mbox{if}& \;\;\;\mbox{$c=0$}\\
(cN)^{-1} &\;\; \mbox{if}& \;\;\;\mbox{$c>0$}\label{N1N}
 \end{array}
 \right. .
 \end{eqnarray}
Note that $N_1/N$ displays the signature of the second order adsorption
transition for $N \to \infty$ with respect to $c$. Therefore, $N_1/N$ plays the role
of the order parameter of the adsorption transition.

The thickness of the adsorbed chain, $\xi_{th}$, is closely related to
the fraction of monomers at the surface $N_1/N$
\cite{EKB82,Eisenriegler}.
 The more monomers are fixed at the wall, the smaller the region
 occupied by the remaining monomers. In particular, for
$ c \leq 0$ one obtains using Eqs.(\ref{c_le_0}) and (\ref{xic})
$N_{1}/N\sim\xi_{th}^{-(1-\Phi)/\nu}$. The first order results of
calculations  for exponents $(1-\Phi)/\Phi$ and $(1-\Phi)/\nu$ are
presented in Table II. The dependence of the fraction of monomers
adsorbed at the surface $N_{1}/N$ on $\xi_{c}$ for different values
$\delta$ for $c\leq 0$ (i.e. below the adsorption threshold) is
presented in Figure 3.

\begin{figure}[htb]
\begin{center}
\includegraphics[scale=0.8]{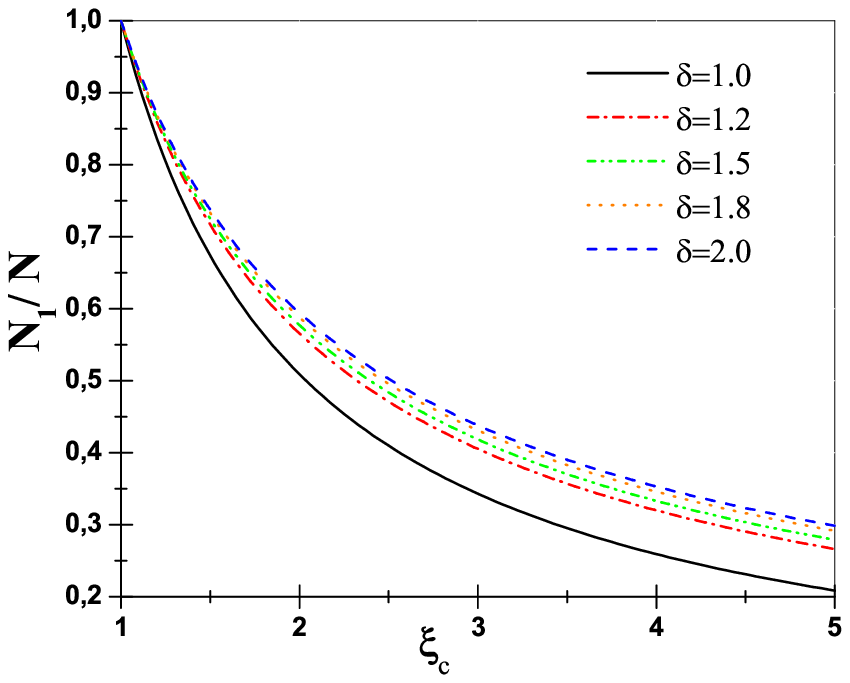}
\end{center}
\caption{The dependence of the fraction of  monomers at the surface
$N_1/N$ on $\xi_{c}$ for different values $\delta$ for $c \leq
0$(i.e. below the  adsorption threshold). Both quantities are
dimensionless.} \label{fig3}
\end{figure}

The knowledge of the above mentioned surface critical exponents give
access to the analysis of the scaling behavior of the mean number of
the free ends in the layer between $z$ and $z+dz$.
 The mean number of the
free ends in the layer between $z$ and $z+dz$ is proportional to the
partition function of a chain with one end fixed at ${\bf
x}_{A}=({\bf r}_{A},z)$ and the other end free, $Z_{N}(z)$, where

\be Z_{N}(z)= \int_{0}^{\infty}dz Z^{\lambda}_{N}(z^{'},z).\ee
Short-distance behavior ($l\ll z\ll \xi_R$) of the $Z_{N}(z)$ right
at the threshold ($c=0$) is \be \label{ZN} Z_{N}(z) \sim z^{a^{'}}
N^{b^{'}} \ee with $a^{'}=(\gamma-\gamma_{1})/ \nu$ and
$b^{'}=\gamma_{1}-1$. Short-distance behavior ($l\ll z\ll \xi_R$) of
the density of monomers in a layer  at the distance $z$ from the
wall to which one end of the polymer is attached,
$M_{N}^{\lambda}(z)$ right at the threshold ($c=0$) is \be
\label{MN} M_{N}^{\lambda}(z)\sim z^{-\bar{a}}N^{\bar{b}}, \ee where
$\bar{a}=1-(1-\Phi)/\nu$, $\bar{b}=-1+\Phi+\gamma_{1}$. The exponent
$\bar{a}$ has been introduced as the proximal exponent by de Gennes
and Pincus~\cite{deGennesP83}. Note that $\Phi > 1-\nu$ is obtained
in all results. Thus, the conjecture by Bouchaud and Vannimenus
\cite{Bouchaud89} is obeyed. This means that the adsorption profile
is strickly decaying with a positive proximal exponent. The
calculation of the dependence of the above mentioned critical
exponents $a^{'},b^{'}$ and $\bar{a},\bar{b}$ from the correlation
parameter $\delta$ is presented in Table II. The scaling behavior of
$Z_{N}(z)$ and $M_{N}^{\lambda}$ for different values of correlation
parameter $\delta$ are presented on the Figure 4 and Figure 5,
respectively.

\begin{figure}[htb]
\begin{center}
\includegraphics[scale=0.8]{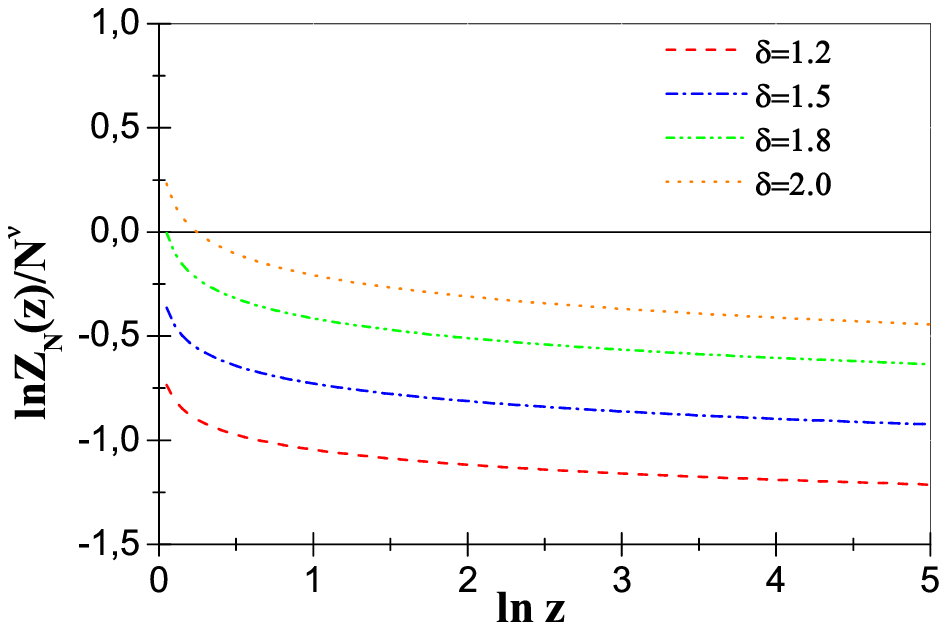}
\end{center}
\caption{The  partition function $ln Z_{N}(z)/N^{\nu}$  just at  the
adsorption threshold $c=0$ and for $N=100$, as a function of $ln z$
for $l\ll z\ll N^{\nu}$ and for different values of $\delta$. }
 \label{fig4}
\end{figure}

\begin{figure}[htb]
\begin{center}
\includegraphics[scale=0.8]{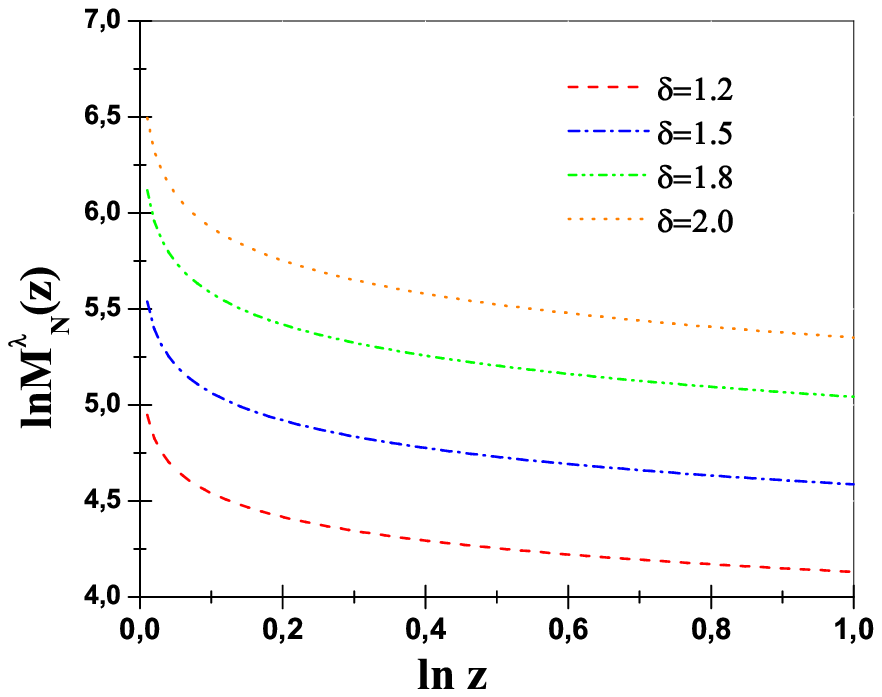}
\end{center}
\caption{The  density of monomers $ln M^{\lambda}_{N}(z)$ in the
layer as function of $ln z$ for $l\ll z \ll N^{\nu}$ just at the
adsorption threshold $c=0$ and for $N=100$ for different values of
 $\delta$. } \label{fig5}
\end{figure}

\section{Conclusions}

In the present paper we have investigated adsorption on "marginal"
and attractive surface of long-flexible polymer chains in media with
quenched long-range correlated surface and decaying near the surface
disorder of the type (\ref{g_def}). The
correlation function given in Eq.(\ref{g_def}) implies also a
layer-like structure of the impurities in the bulk. In such systems
the chains has to go around large correlated regions, and
effectively occupies a large space, with the defects contained
inside the region occupied by the coil and as a result, the polymer
chain swells. This explaines the increase of the
 mean square end-to-end distance and the radius of gyration as a
 result of the increase of the critical
 exponent $\nu$ (see Table I) when the correlation of the disorder is
increased (i.e., $a=3-\delta$ is decreased).

If the range of correlation is very large, then the polymer chain
may be trapped between the walls of defects. From Fig.4 and Fig.5,
Eqs.(\ref{ZN}),(\ref{MN}) and results of Tables I,II, we can see
that the mean number of free ends
 in the layer between $z$ and $z+dz$ and the density of monomers in a
 layer  at the distance $z$ from the wall in the near surface region
 ($l\ll z\ll \xi_R$) both increase for decreasing $a=3-\delta$. From
 the fact that crossover exponent $\Phi$ decreases for decreasing
 $a$ (see Table I) we obtain, that the thickness of the adsorbed layer
 (see Fig.2) and the fraction of monomers $N_{1}/N$ as function of $\xi_{c}$
 increase for decreasing $a$ (see Fig.3). From Eq.(\ref{N1N}) and
 results of Table II we obtain that for small values of $|c|$ the
 fraction of adsorbed monomers $N_{1}/N$ as function of $|c|$
 decreases  when the correlation of the disorder is increased
 (i.e., $a=3-\delta$ is decreased). This indicates that the fraction
 of the monomers near the wall can be higher then directly at the wall when we add disorder.

 Our result in the first
 order of the ($\epsilon,\delta$)-expansion for systems with long range
 correlated surface and decaying near the surface disorder are in
 good agreement with our previous results for systems with
long-range correlated bulk disorder \cite{UC04} (also obtained in the
framework of the one-loop approximation scheme).
We note that the absolute
values for surface critical exponents obtained in the framework of
($\epsilon,\delta$)-expansion are smaller, than those obtained within
the one-loop approximation scheme. A similar situation is found for
pure systems without disorder (see results for
surface critical exponents $\eta_{\parallel}$ and $\eta_{c}$
obtained with the two different schemes \cite{DSh98}).

The obtained results indicate that the long-range correlated surface and
decaying near the surface disorder has influence on the process of
adsorption of
 long-flexible polymer chains on the surface. The system considered in our work
 belongs to a new universality class. All
 sets of surface and bulk  critical exponents depends from correlation parameter
 $\delta$ of the correlation function describing disorder (see
Eq.(\ref{g_def})). As was indicated in \cite{Usat06}, performing of
the further calculations in the framework of two-loop approximation
scheme  directly at fixed dimensions $d=3$ gives possibility to
obtain more reliable quantitative results. It will be the subject of
forthcoming work.

\section*{Appendix 1}
\setcounter{equation}{0}

The general forms of integrals $I_{4}$ and $I_{5}$ for the case
$\xi<<\xi_{R}$ are: \bea
I_{4}&=&\frac{2^{\delta-2}\Gamma[\frac{\delta-\epsilon}{2}](\delta-\epsilon)\pi^{\frac{1-d}{2}}}{
\Gamma[\frac{3-\epsilon}{2}]\Gamma[1-\frac{\delta}{2}]\sin{\frac{\pi\delta}{2}}\Gamma[\frac{\epsilon}{2}]},\nonumber\\
I_{5}&=&\frac{\Gamma[\frac{\delta}{2}]\Gamma[\frac{3-\delta}{2}]}{\Gamma[\frac{3-\epsilon}{2}]
\Gamma[\frac{\epsilon}{2}]}.\label{sd2b} \eea

\section*{Appendix 2}
\setcounter{equation}{0}

The general forms of integrals $I_{4}$ and $I_{5}$ for the case
$\xi\sim\xi_{R}$ are: \be
I_{4}=-\frac{2^{\delta-3}\Gamma[\frac{\delta-\epsilon}{2}](\delta-\epsilon)\pi^{\frac{(1-d)}{2}}}
{(1+\delta)\Gamma[\frac{3-\epsilon}{2}]\sin{\frac{\delta\pi}{2}}\Gamma[-\frac{\delta}{2}]\Gamma[\frac{\epsilon}{2}]},\label{i4kxba}\ee

\be I_{5}=
\frac{2^{\frac{1-\delta}{2}}\delta\sqrt{\pi}\Gamma[\delta]\Gamma[\frac{3-\delta}{2}]}
{2(1+\delta)\Gamma[\frac{3-\epsilon}{2}]\Gamma[\frac{1+\delta}{2}]\Gamma[\frac{\epsilon}{2}]}{}_{2}F_{1}[\frac{5+\delta}{2},\frac{3-\delta}{2},
\frac{5+\delta}{2},\frac{1}{2}].\ee

\section*{Appendix 3}
\setcounter{equation}{0}

The individual RG series expansions for the other critical exponents
can be derived through standard surface scaling relations \cite{D86}
with $d=3$
\begin{eqnarray}
&& \eta_{\perp} = \frac{\eta +
\eta_{\parallel}}{2}, \nonumber\\
&& \beta_{1} = \frac{\nu}{2}
(d-2+\eta_{\parallel}), \nonumber\\
&& \gamma_{11}=\nu(1-\eta_{\parallel}), \nonumber\\
&& \gamma_{1}= \nu(2-\eta_{\perp}), \label{sc}\\
&& \Delta_{1}= \frac{\nu}{2} (d-\eta_{\parallel}), \nonumber\\
&& \delta_{1} = \frac{\Delta}{\beta_{1}} =
\frac{d+2-\eta}{d-2+\eta_{\parallel}}, \nonumber\\
&& \delta_{11} = \frac{\Delta_{1}}{\beta_{1}}=
\frac{d-\eta_{\parallel}}{d-2+\eta_{\parallel}}\;.\nonumber
\end{eqnarray}

Each of these critical exponents characterizes certain properties of
the semi-infinite systems with long-range correlated surface and
decaying near the surface disorder, in the vicinity of the critical
point. The values $\nu$, $\eta$, and $\Delta=\nu(d+2-\eta)/2$ are
the standard bulk exponents.

\section*{Acknowledgments}
This work in part was supported by grant from National Polish
Foundation. Z.U. should like to thanks L.Longa for hospitality at
the Jagiellonian University in Cracow, Poland. We gratefully
acknowledge fruitful discussions with H.W. Diehl and Ch. von Ferber.

\end{document}